\documentclass[aps,pra,onecolumn,superscriptaddress]{revtex4-2}
\usepackage{float}

\usepackage{amssymb,amsmath,amsthm,mathrsfs,amsfonts,dsfont}
\usepackage{bm}
\usepackage{graphicx}
\usepackage{hyperref}
\usepackage{times}
\usepackage{physics}
\usepackage{subfigure}
\usepackage{hyperref} 
\usepackage{color}
\hypersetup{colorlinks=true,citecolor=magenta,linkcolor=magenta,urlcolor=magenta}

\begin{document}


\title{Quantum Annealing with Qubit–Resonator Systems for Simultaneous 
Optimization of\\ Binary and Continuous Variables}


\author{Seiya Endo}
\thanks{These authors equally contributed to this paper.}
\affiliation{Department of Electrical, Electronic, and Communication Engineering, Faculty of Science and Engineering, Chuo University, 1-13-27, Kasuga, Bunkyo-ku, Tokyo 112-8551, Japan}

\author{Shohei Kawakatsu}
\thanks{These authors equally contributed to this paper.}
\affiliation{Department of Electrical, Electronic, and Communication Engineering, Faculty of Science and Engineering, Chuo University, 1-13-27, Kasuga, Bunkyo-ku, Tokyo 112-8551, Japan}

\author{Hiromichi Matsuyama}
\affiliation{Jij Inc., 3-3-6 Shibaura, Minato-ku, Tokyo 108-0023, Japan}

\author{Kohei Suzuki}
\affiliation{Jij Inc., 3-3-6 Shibaura, Minato-ku, Tokyo 108-0023, Japan}

\author{Yuichiro Matsuzaki}
\email[]{matsuzaki.yuichiro@aist.go.jp}
\affiliation{Department of Electrical, Electronic, and Communication Engineering, Faculty of Science and Engineering, Chuo University, 1-13-27, Kasuga, Bunkyo-ku, Tokyo 112-8551, Japan}



\date{\today}

\begin{abstract}
Quantum annealing is a method developed to solve combinatorial optimization problems by utilizing quantum bits. Solving such problems corresponds to minimizing a cost function defined over binary variables. However, in many practical cases, the cost function may also involve continuous variables. Representing continuous variables using quantum bits requires binary encoding, which demands a large number of qubits. To overcome this limitation, an approach using quantum resonators has been proposed, enabling the direct handling of continuous variables within the quantum annealing framework.
{On the other hand}, certain optimization problems involve both binary and continuous variables simultaneously, and a quantum annealing method capable of efficiently solving such hybrid problems has not been established. {Here}, we propose a quantum annealing method based on a hybrid system composed of qubits and resonators, aiming to minimize cost functions that contain both binary and continuous variables. We present a general framework for hybrid quantum annealing using such systems, and {investigate} its feasibility and effectiveness through numerical simulations.
\end{abstract}


\maketitle


%
%
\section{\label{Sec.intro}Introduction}
\label{sec:intro}
Quantum computing technologies based on the principles of quantum mechanics have the potential to efficiently solve optimization problems that are difficult to tackle using classical computers. {One of such technologies} is quantum annealing~\cite{kadowaki1998quantum,farhi2014quantum}. Quantum annealing is a method {to find a ground state of a Hamiltonian, and is used to solve a combinational optimization problem~\cite{santoro2006optimization,hauke2020perspectives}}. In this method, a Hamiltonian representing the problem, {which is called a} problem Hamiltonian, is constructed, and then a driver Hamiltonian representing the transverse field is introduced. The total Hamiltonian of the system is described as follows:
\begin{align}
    \hat{H}(t) &= \left( 1 - \frac{t}{T} \right) \hat{H}_{\mathrm{D}} + \frac{t}{T} \hat{H}_{\mathrm{P}}\label{eq_H}
\end{align}
Here, $t$ denotes time, $T$ is the total time over which quantum annealing is performed, $\hat{H}_{\mathrm{D}}$ is the driver Hamiltonian whose ground state is easy to obtain, and $\hat{H}_{\mathrm{P}}$ is the problem Hamiltonian whose ground state corresponds to the solution.
{The total Hamiltonian is equal to $\hat{H}_{\mathrm{D}}$ ($\hat{H}_{\mathrm{P}}$) at $t=0$ ($t=T$).} 
Through {the dynamics induced by the total Hamiltonian}, 
{the system gradually changes while remaining in the ground state of the total Hamiltonian as long as an adiabatic condition is satisfied}~\cite{hauke2020perspectives,morita2008mathematical,albash2018adiabatic,childs2001robustness,jansen2007bounds}.

Conventional quantum annealing by using qubits has been used to minimize cost functions with binary variables, a method referred to as discrete-variable quantum annealing {in this study}~\cite{kadowaki1998quantum,farhi2014quantum}.
Specifically, such problems are often formulated using the Ising model or the Quadratic Unconstrained Binary Optimization (QUBO) formulation~\cite{kadowaki1998quantum}. Discrete-variable quantum annealing has been implemented in commercial quantum annealers such as the D-Wave machine~\cite{johnson2011quantum,boixo2014evidence,king2022coherent}. On the other hand,  in cases where the cost function is described solely by continuous variables, it is possible to minimize it using qubit-based quantum annealing by introducing auxiliary qubits~\cite{johnson2011quantum}. To convert continuous-variable cost functions into Ising or QUBO forms, an encoding process is necessary, and various techniques such as binary encoding have been proposed~\cite{chancellor2019domain, karimi2019practical, tamura2021performance}. However, {conventional} approaches approximate continuous variables using binary expansions based on binary variables,  {and} increasing the precision requires a greater number of qubits~\cite{date2021adiabatic}.

Recently, quantum annealing using resonators has been proposed to overcome this issue~\cite{koura2024linear}. When using resonators, each mode can provide an infinite-dimensional Hilbert space. This enables the problem Hamiltonian to be constructed using the modes of the resonator, allowing continuous variables to be treated directly rather than being approximately represented using binary variables. This method is referred to as continuous-variable quantum annealing {in this study.} Unlike conventional quantum annealing, this approach can improve precision without increasing the number of quantum devices {as long as an adiabatic condition is satisfied}~\cite{koura2024linear}.

Here, we propose a method using a coupled system of resonators and qubits for solving optimization problems  {whose cost functions include} both continuous and binary variables,  {called} mixed-integer programming {(MIP)}{\cite{khosravi2023mixed}}. 
Binary variables are represented using qubits, and continuous variables are represented using resonators. This approach enables the combined treatment of discrete combinatorial optimization and continuous function optimization, thereby making it applicable to {MIP} that were difficult to handle with conventional methods. Specifically, while conventional quantum annealing requires continuous variables to be approximated by binary expansions, our method enables continuous variables to be represented directly by resonators without approximation. 
After presenting the general framework of our proposed method, we evaluate its performance through numerical simulations. 

The structure of this paper is as follows. Section 2 explains quantum annealing using binary variables. Section 3 describes quantum annealing using continuous variables. Section 4 presents hybrid quantum annealing that handles both binary and continuous variables. Section 5 provides specific numerical calculations for hybrid quantum annealing. Section 6 concludes the paper.

%
%
\section{Theory of Quantum Annealing with Binary Variables}

Quantum annealing is one of the methods used to solve combinatorial optimization problems ~\cite{kadowaki1998quantum,farhi2014quantum}. A combinatorial optimization problem is the task of finding a combination that minimizes the objective function value from among many possible combinations~\cite{santoro2006optimization,hauke2020perspectives}. In quantum annealing, a problem Hamiltonian is constructed such that its ground state corresponds to the optimal solution of an optimization problem described using binary variables. The optimization problem is then solved by searching for the ground state of this problem Hamiltonian. The total Hamiltonian is expressed as in Eq.~\eqref{eq_H}.
Discrete-variable quantum annealing deals with optimization problems
{whose cost function is described as follows:}
\begin{align}
    L_{\mathrm{discrete}}(\sigma) &= \sum_{i = 1}^L {h_i \sigma^{(i)} } + \sum_{i, j = 1}^L {J_{ij} \sigma^{(i)} \sigma^{(j)} }
\end{align}
Here, $\{h_j\}_{j=1}^L$ {corresponds to} an $L$ {real-valued parameters}, and $\{J_{ij}\}_{i,j=1}^L$ {corresponds to} an $L \times L$ {real-valued parameters}.
In most cases, the following driver Hamiltonian is used:
\begin{align}
    \hat{H}_{\mathrm{D}} &=   \sum_{i = 1}^L B_i\hat{\sigma}_x^{(i)}
\end{align}
where $B_i$ represents the strength of the transverse field, and $\hat{\sigma}_x^{{{(i)}}}$ denotes the Pauli-$X$ matrix.

Quantum annealing proceeds through the following steps. First, the ground state of $\hat{H}_{\mathrm{D}}$ is prepared as the initial state. Next, the system evolves in time adiabatically according to $\hat{H}(t)$. By evolving the system adiabatically, it is expected that the final state will be close to the ground state of $\hat{H}_{\mathrm{P}}$. 
{To satisfy the adiabaticity of the evolution, the following condition should be satisfied\cite{hauke2020perspectives, morita2008mathematical, albash2018adiabatic, childs2001robustness} .}
\begin{align}
    \frac{|\bra{E_1(t)} \partial_t \hat{H}(t) \ket{E_0(t)}|}{\left( E_1(t) - E_0(t) \right)^2} \ll 1\label{eqdannetu}
\end{align}
Here, $\ket{E_0(t)}$ and $\ket{E_1(t)}$ denote the instantaneous ground state and first excited state at time $t$, respectively. $E_0(t)$ and $E_1(t)$ represent the ground-state and first excited-state energies at time $t$.
It is worth mentioning that, as the energy gap of the Hamiltonian increases, it becomes easier to satisfy the adiabatic condition.

%
%
\section{Quantum Annealing with Continuous Variables}

In this section, we describe quantum annealing using continuous variables. Continuous-variable quantum annealing enables the minimization of cost functions described by continuous variables by utilizing resonators instead of qubits to represent the problem~\cite{koura2024linear}. In this approach, the modes of the resonators are used to encode the problem, allowing continuous variables to be handled directly. Compared to conventional approaches using qubits, this method can reduce the number of required quantum devices when the cost function involves continuous variables.

Continuous-variable quantum annealing deals with optimization problems
{whose cost function is described as follows.}
\begin{align}
    L_{\mathrm{continuous}}(\theta) &= \sum_{m, l=1, m\neq l}^M\left( \theta_m C_{ml} \theta_l \right) + \sum_{m=1}^M \left( C_{mm} \theta^2_m - D_m \theta_m \right)
\label{eqrenzoku3}
\end{align}
Here, $\{C_{ml}\}_{m,l=1}^M$ corresponds to an $M \times M$  {real-valued parameters}, and $\{D_m\}_{m=1}^M$ corresponds to
an $M$ {real-valued parameters}. The variables $\theta_1$, $\theta_2$, $\cdots$, $\theta_M$ are real values. The cost function in Eq.~\eqref{eqrenzoku3} is quadratic, and the goal is to find the optimal solution that minimizes this 
{multi-variable} quadratic function.

In continuous-variable quantum annealing, the cost function is expressed using creation and annihilation operators, $\hat{a}_m^{\dagger}$ and $\hat{a}_m$, to construct the problem Hamiltonian~\cite{koura2024linear}.
Here, $\hat{a}_m^{\dagger}$ and $\hat{a}_m$ are the creation and annihilation operators for the $m$-th photon mode.
{We adopt the following replacement to construct the problem Hamiltonian from the cost function.}
\begin{align}
    \theta^2_m &\rightarrow \hat{a}^{\dagger}_m \hat{a}_m\\
    \theta_m &\rightarrow \frac{\hat{a}_m + \hat{a}_m^{\dagger}}{2}\\
    \theta_m \theta_l &\rightarrow \frac{\hat{a}_m \hat{a}_l^{\dagger}}{2} + \frac{\hat{a}_m^{\dagger} \hat{a}_l}{2}
\end{align}

The following driver Hamiltonian is chosen:
\begin{align}
    \hat{H}_{\mathrm{D}} = \sum_{m = 1}^M \omega_{\mathrm{D}}^{(m)} \hat{a}_m^{\dagger} \hat{a}_m
\end{align}
where $\omega_{\mathrm{D}}^{(m)}$ is the frequency of the $m$-th resonator. This Hamiltonian has a trivial ground state, namely the vacuum state. 
{In the continuous {-} variable quantum annealing,}
the system is initialized in the ground state of $\hat{H}_{\mathrm{D}}$ and evolved adiabatically to the ground state of $\hat{H}_{\mathrm{P}}$.
After time evolution, the value of $\theta_m$ is obtained by measuring the expectation value of $\frac{\hat{a}_m + \hat{a}_m^{\dagger}}{2}$. Since this operator corresponds to the amplitude of a coherent state, if the system evolves into a coherent state during quantum annealing, the amplitude can take continuous values. Therefore, unlike discrete-variable quantum annealing, which requires approximating continuous variables using binary ones, continuous-variable quantum annealing allows direct treatment of continuous variables.

%
%
\section{Theory of Hybrid Quantum Annealing with Binary and Continuous Variables}

In discrete-variable and continuous-variable quantum annealing, only binary or continuous variables can be treated, respectively. However, in practice, there exist optimization problems that involve both binary and continuous variables, such as {MIP} .

To address such problems, we propose a  quantum annealing approach that combines discrete-variable and continuous-variable quantum annealing. {This method is referred to as hybrid quantum annealing in this study.} In the proposed method, binary variables are represented using qubits, while continuous variables are represented using resonator modes. This approach enables the direct handling of both binary and continuous variables, allowing {MIP}   to be addressed without approximation~\cite{koura2024linear}.
Hybrid quantum annealing targets optimization problems
{whose cost function is described as follows.}
\begin{align}
    L(\theta) &= L_{\mathrm{continuous}}(\theta) +L_{\mathrm{discrete}}(\sigma) + \sum_{i = 1}^L \sum_{j = 1}^M { \left( g_{ij} \sigma^{(i)} \theta_j^2 + 2 \tilde{g}_{ij} \sigma^{(i)} \theta_j \right) }\label{eqref:costkansuu}
\end{align}
Here, $g{ij}$ and $\tilde{g}_{ij}$ are $L \times M$ matrices. The variables $\theta_1$, $\theta_2$, $\cdots$, $\theta_M$ are real-valued, and $\sigma^{(1)}$, $\sigma^{(2)}$, $\cdots$, $\sigma^{(L)}$ are binary variables taking values $\pm1$.
In this formulation, $L_{\mathrm{discrete}}$ typically corresponds to a {QUBO} form, and $L_{\mathrm{continuous}}$ is a function dependent on continuous variables. The third term in Eq.\eqref{eqref:costkansuu} involves interactions between binary and continuous variables\cite{das2005quantum}.
The binary variables $y_i$ and continuous variables $x_i$ are mapped to quantum operators as follows:
\begin{align}
    x_i &\rightarrow \frac{1}{2} \left( \hat{a}_i + \hat{a}_i^{\dagger} \right)\label{eq_xi}\\
    x_i^2 &\rightarrow  \hat{a}_i^{\dagger} \hat{a}_i\label{eq_xi^2}\\
    y_i &\rightarrow \frac{\hat{\mathbf{1}} + \hat{\sigma}_z ^{(i)}}{2}\label{eq_yi}
\end{align}

The problem Hamiltonian corresponding to this cost function is expressed as:

\begin{align}
    \hat{H}_{\mathrm{P}} = \hat{H}_{\mathrm{discrete}} + \hat{H}_{\mathrm{continuous}} + \hat{H}_{\mathrm{interaction}}\label{eqref:risantorenzokuhp}
\end{align}
Here, $\hat{H}_{\mathrm{discrete}}$ corresponds to the binary-variable part, $\hat{H}_{\mathrm{continuous}}$ to the continuous-variable part, and $\hat{H}_{\mathrm{interaction}}$ to the interaction between them. More explicitly, Eq.~\eqref{eqref:risantorenzokuhp} is written as:

\begin{align}
    \hat{H}_{\mathrm{discrete}} &= \sum_{i = 1}^L {h_i \hat{\sigma}_z^{(i)}} + \sum_{i<j}^L {J_{ij} \hat{\sigma}_z^{(i)}\hat{\sigma}_z^{(j)}}
\end{align}
\begin{align}
    \hat{H}_{\mathrm{interaction}} &= \sum_{i = 1}^L \sum_{j = 1}^M \left( g_{ij} \hat{\sigma}_z^{(i)} \hat{a}_j^{\dagger} \hat{a}_j + \tilde{g}_{ij} \hat{\sigma}_z^{(i)} \left( \hat{a}_j + \hat{a}^{\dagger}_j \right) \right)
\end{align}
\begin{align}
    \hat{H}_{\mathrm{continuous}} &= \sum_{i = 1}^M {
    \left(
    \omega_{\mathrm{c}}^{(i)} \hat{a}^{\dagger}_i \hat{a}_i + \lambda_i \left( \hat{a}_i + \hat{a}^{\dagger}_i \right) \right)+ \sum_{i<j}^{M} { \tilde{J}_{ij} \left( \hat{a}_i \hat{a}_j^{\dagger} + \hat{a}_i^{\dagger} \hat{a}_j \right)  } }
\end{align}
Here, $h_j$ is the resonance frequency of the $j$-th qubit, $J_{ij}$ represents the interaction between the $i$-th and $j$-th qubits, $g_{ij}$ is the coupling strength between the $i$-th qubit and the $j$-th resonator, and $\hat{a}_j$, $\hat{a}_j^{\dagger}$ are the annihilation and creation operators of the $j$-th resonator. $\tilde{g}_{ij}$ denotes the coupling between the $i$-th qubit and the $j$-th resonator, $\omega_{\mathrm{c}}^{(i)}$ is the frequency of the $i$-th resonator, $\lambda_j$ is the external field strength applied to the $j$-th resonator, and $\tilde{J}_{ij}$ is the interaction between the $i$-th and $j$-th resonators.
The corresponding driver Hamiltonian is expressed as:
\begin{align}
    \hat{H}_{\mathrm{D}} = \sum_{i = 1}^M \omega_{\mathrm{c}}^{(i)} \hat{a}_i^{\dagger} \hat{a}_i + \sum_{j = 1}^L B_{{j}}\hat{\sigma}_x^{({j})}\label{eqref:risantorenzokuhd}
\end{align}
where \(B_{i}\) denotes the strength of the transverse field applied to the $i$-th qubit.
In our method, the ground state of the driver Hamiltonian $\hat{H}_{\mathrm{D}}$ in Eq.\eqref{eqref:risantorenzokuhd} is prepared as the initial state. By performing adiabatic time evolution according to Eq.\eqref{eq_H}, the ground state of the problem Hamiltonian can be obtained, yielding the minimum of the cost function in Eq.~\eqref{eqref:costkansuu}.

However, in order to solve a problem using {hybrid} quantum annealing where the cost function has a nontrivial solution, it is necessary to construct a problem Hamiltonian that also has a nontrivial ground state. To achieve this, the interaction strengths and the resonance frequencies of the Hamiltonian must be on the same order~\cite{kadowaki1998quantum,johnson2011quantum}. In practice, however, when using real quantum hardware, the coupling constants between qubits and resonators are typically several orders of magnitude smaller than the resonance frequencies~\cite{wallraff2004strong,blais2004cavity}, making {such an} implementation difficult.

{Fortunately, we can overcome such a problem by using spin-lock-based quantum annealing \cite{chen2011experimental,matsuzaki2020quantum,imoto2022obtaining}. In the Appendix, we explain
}
 a method to enhance the effective interaction strengths by driving the system with a microwave (oscillating magnetic field) and moving to a rotating frame.
{Due to this mechanism, our proposal would be feasible even with the current technology.}

{On the other hand,} a method for solving {MIP}  using quantum computers based on resonators has been proposed~\cite{khosravi2023mixed}. However, this approach requires, for example, an infinitely strong squeezing for the initial state  {and four-body interaction}.  These could make experimental implementation  challenging,
which is a contrast to our proposed method {which} has the advantage of using experimentally accessible initial states and interaction types.

%
%

\section{Numerical Results of Hybrid Quantum Annealing with Binary and Continuous Variables}

In this section, we show that small-scale {MIP}  can be solved using hybrid quantum annealing through numerical simulations. As an example of a mixed-integer programming problem, we consider a simplified small-scale production planning problem.

We consider a production scenario in which a single product is manufactured across $K$ production lines, and aim to minimize the total production cost. The cost of producing an amount $x_i \in \mathbb{R}$ on production line $i$ is assumed to grow approximately linearly {when the cost is small, but grow quadratically beyond some point.}
The production cost for producing $x_i$ units on line $i$ is modeled as $c_i x_i + d_i x_i^2$ where $c_i$ is the linear coefficient representing the unit production cost on line $i$ and $d_i$ reflects the degree of nonlinearity in the cost per unit.
We further assume that an additional capital investment of $b_i y_i$ can be made in line $i$, which reduces the linear cost term by $\tilde{c}_i y_i$. Here, $b_i$ denotes the investment cost, and $\tilde{c}_i$ represents the cost reduction effect per unit when the investment is made. The objective is to minimize the total cost while satisfying a required total production quantity.

{Here, $x_i$ is a continuous variable representing the production quantity on production line $i$, and $y_i$ is a binary variable indicating whether to invest in production line $i$.}
The objective function is given by:
\begin{align}
    &\underset{x, y}{\mathrm{min}} \sum_{i=1}^{K} {c_i x_i} - \sum_{i=1}^{K} {\tilde{c}_i y_i x_i} + \sum_{i=1}^{K} {d_i {x_i}^2} + \sum_{i=1}^{K} {b_i y_i} \notag\\
    &= \underset{x, y}{\mathrm{min}} \sum_{i=1}^{K} \left\{ {\left( c_i - \tilde{c}_i y_i \right) x_i}  \right\} + \sum_{i=1}^{K} {d_i {x_i}^2}  + \sum_{i=1}^{K} {b_i y_i}\label{eq_mokuteki}
\end{align}
To ensure that the total production quantity equals $A$, we impose the following constraint:
\begin{align}
    \sum_i^K {x_i} &= A\label{eq_seiyaku}
\end{align}

The resulting optimization problem can be formulated as follows:

\begin{align}
    \underset{x, y}{\mathrm{min}} \sum_{i=1}^{K} \left\{ {\left( c_i - \tilde{c}_i y_i \right) x_i}  \right\} &+ \sum_{i=1}^{K} {d_i {x_i}^2}  + \sum_{i=1}^{K} {b_i y_i}\\
    \text{s.t.} \sum _i ^K {x_i} &= A\\
    x_i &\geq 0 \\
    x_i \in \mathbb{R}, \ & y_i \in \{ 0, 1 \}
\end{align}
By applying the penalty method to relax the constraint, we obtain the following cost function:
\begin{align}
    \underset{x, y}{\mathrm{min}} \left( \sum _i ^K {\left( c_i - \tilde{c}_i y_i \right) x_i } + \sum _i ^K {d_i x^2_i} + \sum_i ^K { b_i y_i } + \lambda \left( \sum _i ^K {x_i} - A \right)^2  \right)\label{eq_cost}
\end{align}
Substituting Eqs.\eqref{eq_xi}, \eqref{eq_xi^2}, and \eqref{eq_yi} into Eq.\eqref{eq_cost}, the corresponding problem Hamiltonian $H_{\mathrm{P}}$ is given by:
\begin{align}
    \hat{H}_{\mathrm{P}} &= \sum _i ^K { \left( \left( \hat{\mathbf{1}}\  c_i - \tilde{c}_i \frac{\hat{\mathbf{1}} + \hat{\sigma}_z ^{(i)}}{2} \right) \frac{1}{2} \left( \hat{a}_i + \hat{a}_i^{\dagger} \right) \right)} + \sum _i ^K \left(d_i \hat{a}_i^{\dagger} \hat{a}_i \right) \notag\\
    &+ \sum_i ^K \left( b_i \frac{\hat{\mathbf{1}} + \hat{\sigma}_z ^{(i)}}{2} \right) + \lambda \left( \sum _i ^K {\frac{1}{2} \left( \hat{a}_i + \hat{a}_i^{\dagger} \right)} - A \right)^2  \label{eq_Hp_K}
\end{align}
The corresponding driver Hamiltonian $\hat{H}_{\mathrm{D}}$ is defined as:
\begin{align}
    \hat{H}_{\mathrm{D}} &= \sum_{i = 1}^K \left( {\frac{B_i}{2} \hat{\sigma}_x^{(i)} + {\omega_i \hat{a}^{\dagger}_i \hat{a}_i} } \right)
\end{align}

\begin{figure}[htbp]
    \centering
    \includegraphics[width =1.1\linewidth]{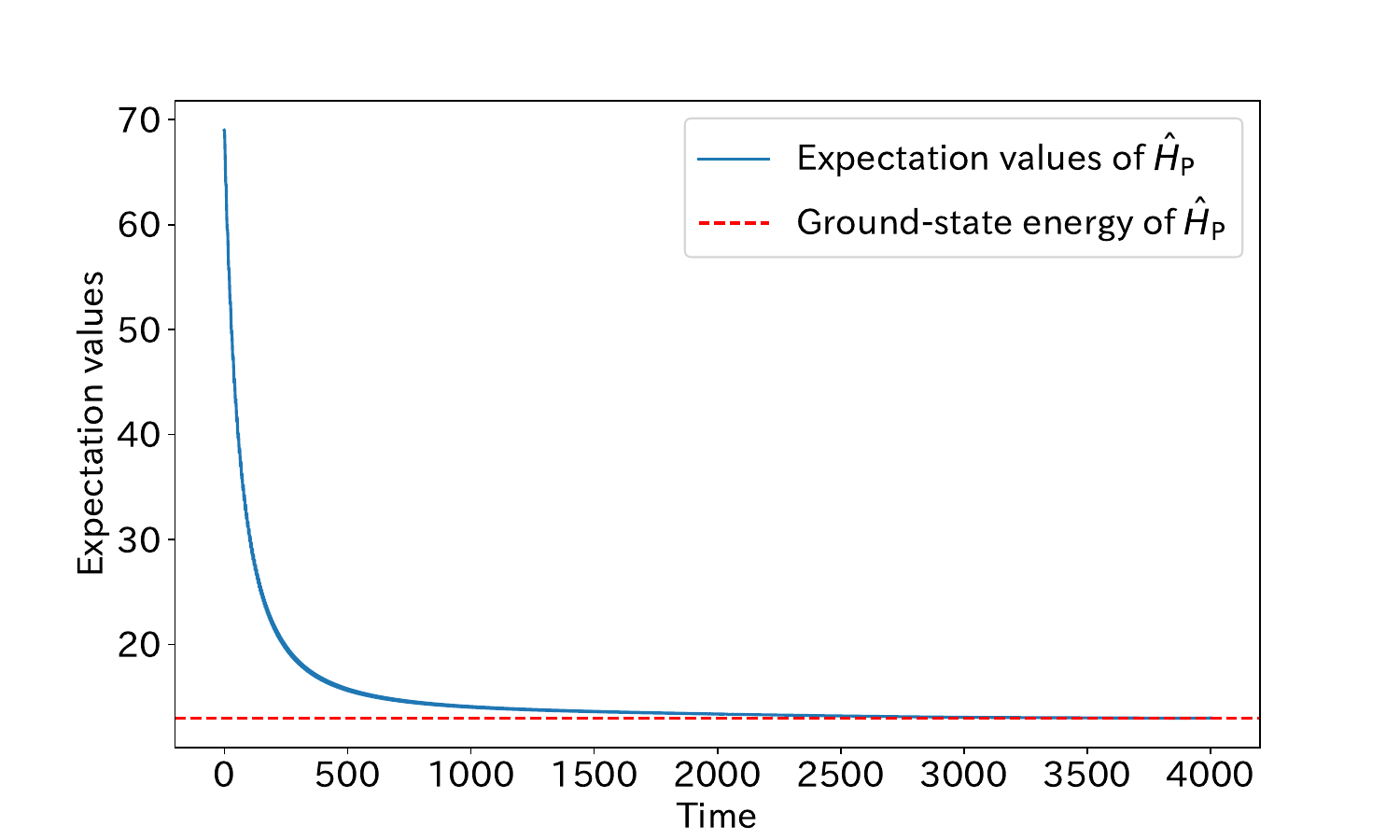}
    \caption{{The time evolution of the expectation values of $\hat{H}_{\mathrm{P}}$}. The blue solid line shows the time evolution of the expectation value of \( \hat{H}_{\mathrm{P}} \), while the red dashed line indicates the ground-state energy of \( \hat{H}_{\mathrm{P}} \). The parameters are set as 
    \( T = 4000.0, \ A  = 2,\ b_1 = 1,\ b_2 = 2,\ c_1 = 2.1,\ c_2 = 2.2,\ \tilde{c}_1 = 1.8,\ \tilde{c}_2 = 2.0,\ d_1 = 3.3,\ d_2 = 3.8,\ \omega_1 = 1.0,\ \omega_2 = 1.0,\ \lambda = 15,\ ,\ B_1 = 1.0,\ B_2 = 1.0,\ K = 2 \).}
    \label{fig_hp}
\end{figure}

\begin{figure}[H]
    \centering
    \includegraphics[width=1\linewidth]{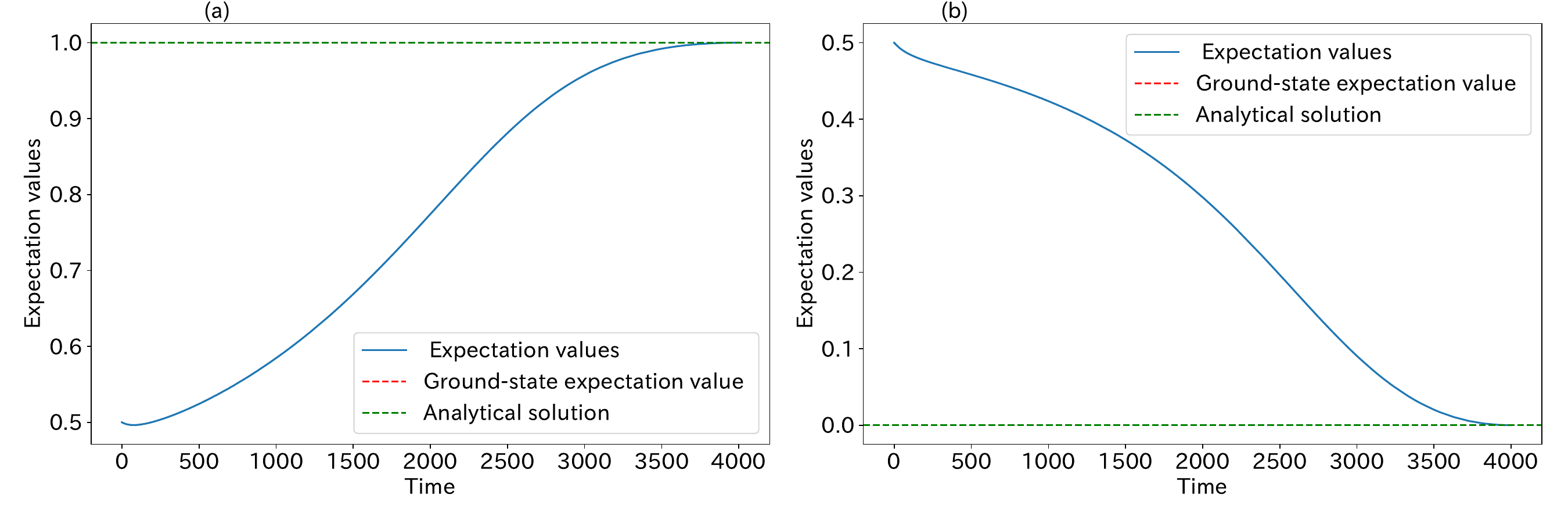}
    \caption{{The time evolution of the expectation values of $\frac{\hat{\mathbf{1}} + \hat{\sigma}_z^{(i)}}{2}$}. The blue solid lines show the time evolution of the expectation values of \( \frac{\hat{\mathbf{1}} + \hat{\sigma}_z^{(1)}}{2} \) and \( \frac{\hat{\mathbf{1}} + \hat{\sigma}_z^{(2)}}{2} \), 
    for (a) and (b),
    respectively. These operators correspond to the binary decision variables \( y_i \), where \( y_i = 0 \) indicates no investment and \( y_i = 1 \) indicates investment in the \( i \)th production line. The red dashed lines show the expectation values of the same operators in the ground state of \( \hat{H}_{\mathrm{P}} \), and the green dashed lines show the analytical solutions that minimize the cost function. The parameters used are the same as in Fig.~\ref{fig_hp}.}
    \label{fig_y1y2}
\end{figure}

\begin{figure}[h]
    \centering
    \includegraphics[width=1\linewidth]{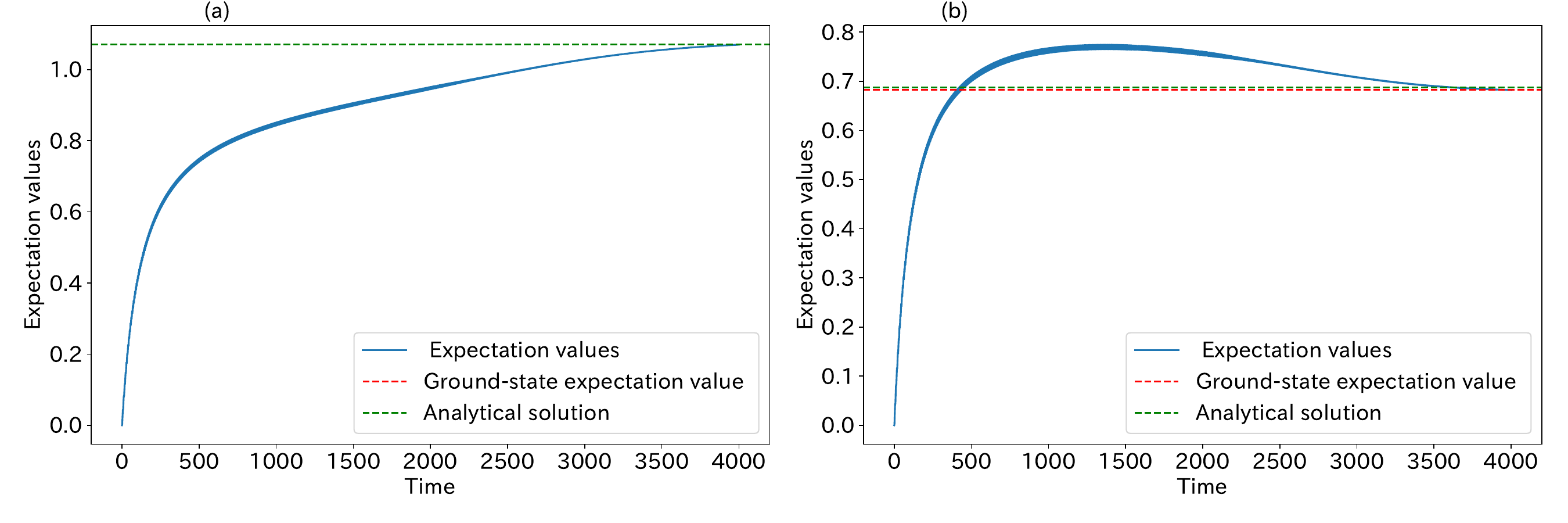}
    \caption{ {The time evolution of the expectation values of $\frac{1}{2} \left( \hat{a}_i + \hat{a}_i^{\dagger} \right)$}. The blue solid lines show the time evolution of the expectation values of \( \frac{1}{2} \left( \hat{a}_1 + \hat{a}_1^{\dagger} \right) \) and \( \frac{1}{2} \left( \hat{a}_2 + \hat{a}_2^{\dagger} \right) \), for (a) and (b), respectively. The red dashed lines indicate the expectation values of these operators in the ground state of \( \hat{H}_{\mathrm{P}} \), while the green dashed lines represent the analytical solutions that minimize the cost function. The parameters used are the same as in Fig.~\ref{fig_hp}.}
    \label{fig_x1x2}
\end{figure}

In this study, we consider the case $K = 2$. In this case, the problem corresponds to two qubits and two resonators, and the driver Hamiltonian $\hat{H}_{\mathrm{D}}$ is given by Eq.~\eqref{eq_Hd}:
\begin{align}
    \hat{H}_{\mathrm{D}} &= \frac{B_1}{2} \sigma_x^{(1)} + \frac{B_2}{2} \sigma_x^{(2)} + \omega_1 \hat{a}_1^{\dagger} \hat{a}_1+ \omega_2 \hat{a}_2^{\dagger} \hat{a}_2\label{eq_Hd}
\end{align}
The corresponding problem Hamiltonian $\hat{H}_\mathrm{P}$ is expressed as follows:
\begin{align}
    \hat{H}_{\mathrm{P}} &=  { \left( \left( \hat{\mathbf{1}}\ c_1 - \tilde{c}_1 \frac{\hat{\mathbf{1}} + \hat{\sigma}_z ^{(1)}}{2} \right) \frac{1}{2} \left( \hat{a}_1 + \hat{a}_1^{\dagger} \right) \right)} + { \left( \left( \hat{\mathbf{1}}\ c_2 - \tilde{c}_2 \frac{\hat{\mathbf{1}} + \hat{\sigma}_z ^{(2)}}{2} \right) \frac{1}{2} \left( \hat{a}_2 + \hat{a}_2^{\dagger} \right) \right)} \notag\\
    &+  \left(d_1 \hat{a}_1^{\dagger} \hat{a}_1 \right) + \left(d_2 \hat{a}_2^{\dagger} \hat{a}_2 \right) +  \left( b_1 \frac{\hat{\mathbf{1}} + \hat{\sigma}_z ^{(1)}}{2} \right) + \left( b_2 \frac{\hat{\mathbf{1}} + \hat{\sigma}_z ^{(2)}}{2} \right) + \lambda \left(  {\frac{1}{2} \left( \hat{a}_1 + \hat{a}_1^{\dagger} \right)} + {\frac{1}{2} \left( \hat{a}_2 + \hat{a}_2^{\dagger} \right)} - A \right)^2  \label{eq_Hp}
\end{align}
By solving the time-dependent Schrödinger equation using Eqs.\eqref{eq_Hd} and \eqref{eq_Hp}, we simulate the dynamics of the quantum annealing process. The results are shown in Figs.\ref{fig_hp}, \ref{fig_y1y2}, and \ref{fig_x1x2}. The parameters used in the simulation are:
$T = 4000.0, \ A  = 2,\ b_1 = 1,\ b_2 = 2,\ c_1 = 2.1,\ c_2 = 2.2,\ \tilde{c}_1 = 1.8,\ \tilde{c}_2 = 2.0,\ d_1 = 3.3,\ d_2 = 3.8,\ \omega_1 = 1.0,\ \omega_2 = 1.0,\ \lambda = 15,\  B_1 = 1.0,\ B_2 = 1.0,\ K = 2$.
\newpage

The analytical solution of this problem is given by 
$y_1 = 1,\ y_2 = 0$, and $x_1 = 1.07,\ x_2 = 0.69$ .
From Fig.~\ref{fig_hp}, we observe that the expectation value of $\hat{H}_{\mathrm{P}}$, obtained by solving the time-dependent Schrödinger equation through {hybrid} quantum annealing, coincides with the ground-state energy of the problem Hamiltonian. This confirms that the ground state has been successfully obtained by {hybrid} quantum annealing.
Furthermore, in Figs.~\ref{fig_y1y2} and \ref{fig_x1x2}, the analytically computed optimal discrete variables $y_i$ and continuous variables $x_i$ , which minimize the cost function, are shown as green dashed lines. These values are consistent with the expectation values of $\frac{\hat{\mathbf{1}} + \hat{\sigma}_z ^{(i)}}{2}$ and $\frac{1}{2} \left( \hat{a}_i + \hat{a}_i^{\dagger} \right)$, respectively, obtained from the {hybrid} quantum annealing result.
Therefore, we have successfully solved a small-scale mixed-integer programming problem using numerical simulation of {hybrid} quantum annealing.

%
%
\section{Conclusion} \label{chap:conclusion}

{Here,} we propose a new framework for quantum annealing, hybrid quantum annealing, which allows the simultaneous treatment of continuous and binary variables. In the proposed method, binary variables are represented using qubits, while continuous variables are encoded in the modes of quantum resonators. This allows for the direct solution of {MIP}  using quantum annealing.
To demonstrate the effectiveness of our approach, we applied it to a small-scale production planning problem as an example of a mixed-integer programming task. Through numerical simulations, we confirm that the proposed method successfully finds the optimal solution to this problem.
These results suggest new possibilities for solving optimization problems using quantum annealing. Future work will focus on extending this method to larger-scale systems, which is expected to further advance its practical applicability.
%
%

\section*{Acknowledgments}

The authors would like to express their sincere gratitude to the developers of QuTiP, which was used for the numerical simulations in this study~\cite{johansson2012qutip}.
The authors thank Asuka Koura for the helpful discussion.
This work is supported by JST Moonshot (Grant Number JPMJMS226C), CREST (JPMJCR23I5), and Presto JST (JPMJPR245B).
This work was performed for Council for Science, Technology and Innovation (CSTI), Cross-ministerial Strategic Innovation Promotion Program (SIP), ``Promoting the application of advanced quantum technology platforms to social issues’' (Funding agency: QST).

\appendix
\section{Hybrid quantum annealing with oscillating magnetic fields}
Here, we explain hybrid quantum annealing with oscillating magnetic fields.
In practical quantum devices, the interaction strength between resonators and qubits is typically several orders of magnitude smaller than the resonance frequency~\cite{wallraff2004strong,blais2004cavity}. As a result, it is challenging to construct a problem Hamiltonian with a nontrivial ground state in hybrid quantum annealing systems composed of solid-state qubits and resonators.
To overcome this problem, we describe a method for realizing hybrid quantum annealing using qubits and resonator modes by effectively enhancing the interaction strength. This is achieved by driving the coupled qubit–resonator system with microwaves and moving into a rotating frame, which relatively amplifies the interaction strength in an effective Hamiltonian representation.

\subsection{General Framework of the Proposed Method}
We begin by rewriting Eq.~\eqref{eqref:risantorenzokuhp} as a modified problem Hamiltonian \(\hat{H}_{\mathrm{P}}^{\prime}\), in which an oscillating magnetic field with frequency \( \omega \) is applied.
\begin{equation}
    \begin{aligned}
        \hat{H}_{\mathrm{P}}^{\prime} =&\sum_{i=1}^{L} \frac{h_i}{2}\hat{\sigma}_z^{(i)}
        +
        \sum_{i<j}^{L} J_{ij}\hat{\sigma}_z^{(i)}\hat{\sigma}_z^{(j)}
        +\sum_{i=1}^{L} \sum_{j=1}^{M} \left(
        g_{ij}\hat{\sigma}_z^{(i)}\hat{a}^\dagger_{j}\hat{a}_{j}
        +
         \tilde{g}_{ij}\hat{\sigma}_z^{(i)}(\hat{a}_{j}+\hat{a}^\dagger_{j})\cos{\omega t}
        \right)
        \\
        +&\sum_{i=1}^{M}
        \left(
        \omega_{\mathrm{c}}^{(i)} \hat{a}^\dagger_{i}\hat{a}_{i}
        + 
        \lambda_i (\hat{a}_{i}+\hat{a}^\dagger_{i})\cos{\omega t}
        \right)
        +
        \sum_{i<j}^{M}\tilde{J}_{i,j}\left(
        \hat{a}_{i}\hat{a}_{j}^\dagger + \hat{a}_{i}^\dagger\hat{a}_{j}\right)
    \end{aligned}
\end{equation}
We also introduce \(\hat{H}_{\mathrm{D}}^{\prime}\) and $\hat{H}^{\prime}(t)$, which are defined as follows.
\begin{equation}
    \hat{H}_{\mathrm{D}}^{\prime} = \sum_{i=1}^{L}
    \left(
    \frac{h_i}{2}\hat{\sigma}_z^{(i)}
    +
    B_i\hat{\sigma}_x^{(i)}\cos{\omega t}
    \right)
    +
    \sum_{j=1}^{M}\omega_{\mathrm{c}}^{(j)} \hat{a}^\dagger_{j}\hat{a}_{j}
\end{equation}
\begin{align}
    \hat{H}^{\prime}(t) &= \left( 1 - \frac{t}{T} \right) \hat{H}_{\mathrm{D}}^{\prime} + \frac{t}{T} \hat{H}_{\mathrm{P}}^{\prime}
    \label{eq_H_dash}  
\end{align}
Furthermore, we introduce the time evolution operator \(\hat{U}\) defined as
\begin{equation}
    \hat{U} = \exp\left(
    \sum_{k=1}^{L} \frac{i\omega t}{2}\hat{\sigma}_z^{(k)}
    +
    \sum_{j=1}^{M} i\omega t \hat{a}^\dagger_{j}\hat{a}_{j}
    \right)
\end{equation}
By transforming \(\hat{H}_{\mathrm{P}}^{\prime}\) and \(\hat{H}_{\mathrm{D}}^{\prime}\) into the rotating frame using this operator and applying the rotating wave approximation (RWA), we obtain the following effective Hamiltonians.

\begin{equation}
    \begin{aligned}
            \hat{H}_{\mathrm{P}}^{\mathrm{\mathrm{eff}}} &= \hat{U}\hat{H}_{\mathrm{P}}^{\prime}\hat{U}^\dagger + i\frac{\partial \hat{U}}{\partial t}\hat{U}^\dagger
            \\
            &\simeq\sum_{i=1}^{L} \frac{h_i-\omega}{2}\hat{\sigma}_z^{(i)}
        +
        \sum_{i<j}^{L} J_{ij}\hat{\sigma}_z^{(i)}\hat{\sigma}_z^{(j)}
        +\sum_{i=1}^{L} \sum_{j=1}^{M} \left(
        g_{ij}\hat{\sigma}_z^{(i)}\hat{a}^\dagger_{j}\hat{a}_{j}
        +
        \frac{ \tilde{g}_{ij}}{2}\hat{\sigma}_z^{(i)}(\hat{a}_{j}+\hat{a}^\dagger_{j})
        \right)
        \\
        &+\sum_{i=1}^{M}(\omega_{\mathrm{c}}^{(i)}-\omega) \hat{a}^\dagger_{i}\hat{a}_{i}
        + 
        \sum_{i=1}^{M} \frac{\lambda_i}{2} (\hat{a}_{i}+\hat{a}^\dagger_{i})
        +
        \sum_{i<j}^{M}\tilde{J}_{i,j}\left(
        \hat{a}_{i}\hat{a}_{j}^\dagger + \hat{a}_{i}^\dagger\hat{a}_{j}\right)
        \label{ippankei.7}
    \end{aligned}
\end{equation}
\begin{equation}
    \begin{aligned}
         \hat{H}_{\mathrm{D}}^{\mathrm{eff}} &= \hat{U}\hat{H}_{\mathrm{D}}^{\prime}\hat{U}^\dagger + i\frac{\partial \hat{U}}{\partial t}\hat{U}^\dagger
            \\
            &\simeq\sum_{i=1}^{L} \frac{h_i-\omega}{2}\hat{\sigma}_z^{(i)}
    +
    \sum_{i=1}^{L} \frac{B_i}{2}\hat{\sigma}_x^{(i)}
    +
    \sum_{j=1}^{M}(\omega_{\mathrm{c}}^{(j)}-\omega) \hat{a}^\dagger_{j}\hat{a}_{j}
    \label{ippankei.8}
    \end{aligned}
\end{equation}
Using Eqs.~\eqref{ippankei.7} and \eqref{ippankei.8}, we construct the full effective Hamiltonian \(\hat{H}^{\mathrm{eff}}(t)\) as follows.
\begin{equation}
    \hat{H}^{\mathrm{eff}}(t)=\left( 1-\dfrac{t}{T}\right) \hat{H}_{\mathrm{D}}^{\mathrm{eff}} + \left(\dfrac{t}{T}\right)\hat{H}_{\mathrm{P}}^{\mathrm{eff}}
    \label{heff(t)}
\end{equation}
Then, by using the ground state of Eq.\eqref{ippankei.8}, which is a trivial solution, as the initial state, we solve the Schrödinger equation under the time-dependent Hamiltonian \(\hat{H}^{\mathrm{eff}}(t)\). Through this time evolution, quantum annealing allows us to obtain the nontrivial solution corresponding to the ground state of Eq.\eqref{ippankei.7}.

{Let us explain a practical advantage of using the oscillating magnetic field for hybrid quantum annealing.}
When the energies of the resonators and qubits are much larger than the interaction strength, only trivial ground states can be obtained, which prevents the practical implementation of quantum annealing .
To address this issue, in our method we appropriately choose the frequency \(\omega\) of the applied oscillating magnetic field, as in Eq.~\eqref{ippankei.7}, so that the relative energies of the resonators and qubits are effectively lowered. This allows us to satisfy the condition:
\begin{equation}
    \frac{\left(
    h_i-\omega
    \right)}{2},
    \left(
    \omega_{\mathrm{c}}^{(i)}-\omega
    \right)
    \simeq
    g_{ij},\frac{ \tilde{g}_{ij}}{2}
\end{equation}
As a result, even in systems where the coupling constants are several orders of magnitude smaller than the resonator frequencies, it becomes possible to prepare nontrivial ground states by executing quantum annealing.
\subsection{Numerical Results}
\label{sec:a+adag3}
\begin{figure}[h!t]
    \centering
    \includegraphics[width=\linewidth]
    {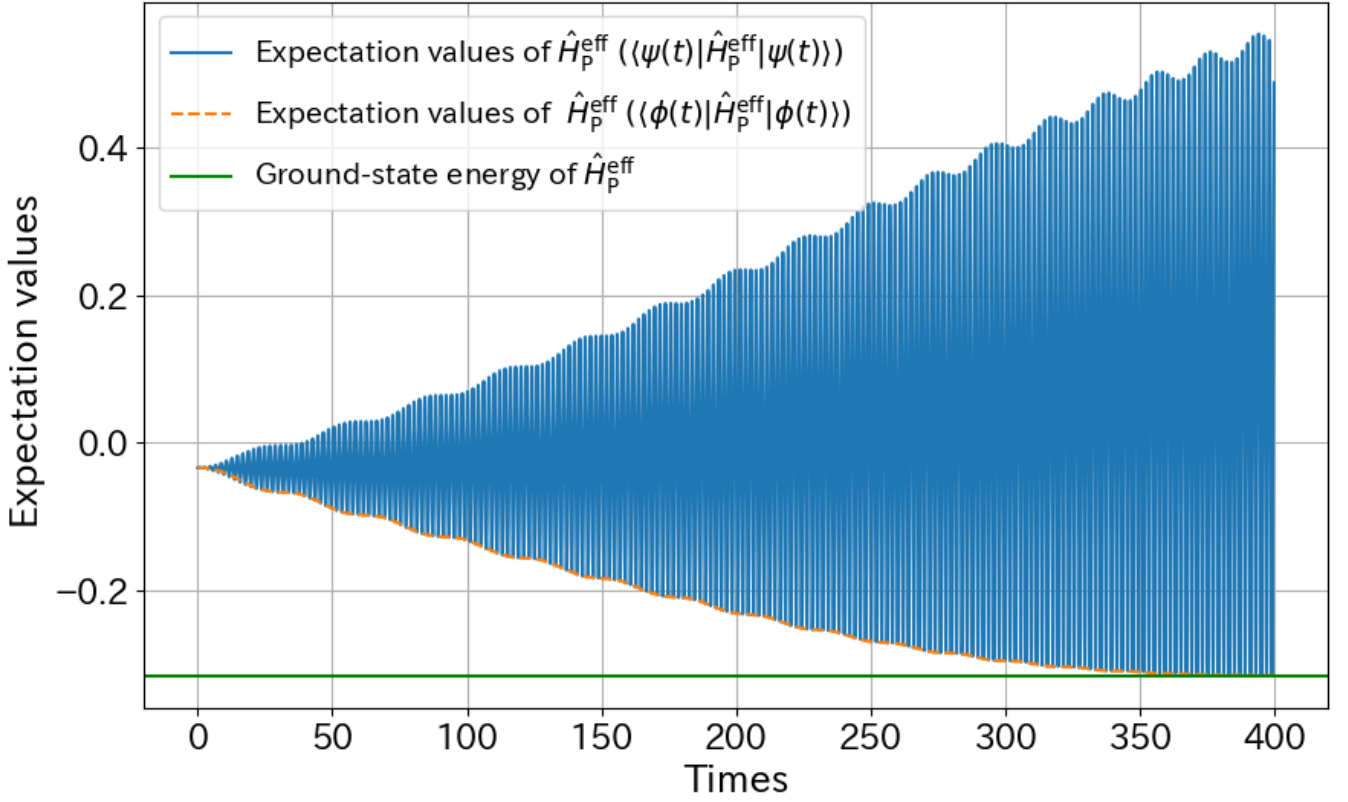}
    \caption{Result of solving the Schrödinger equation for a system in which a single resonator is coupled to a single qubit.  
    The expectation value of \( \hat{H}_{\mathrm{P}}^{\mathrm{eff}} \) at each time \( t \) is plotted.  
    The parameters used are: \( T = 400,\; \omega = 153.9,\; h = 153.7,\; \omega_{\mathrm{c}} = 154.1,\; B = 0.55,\; \lambda = 0.30,\; g = 0.15,\; \tilde{g} = 0.25\).  
    The vertical axis represents the expectation value of \( \hat{H}_{\mathrm{P}}^{\mathrm{eff}} \), and the horizontal axis represents time \( t \).  
    The blue line shows \( \langle \psi(t) | \hat{H}_{\mathrm{P}}^{\mathrm{eff}} | \psi(t) \rangle \), where the initial state \( \ket{\psi(0)} \) is the ground state of \( \hat{H}_{\mathrm{D}}^{\mathrm{eff}} \), and time evolution is performed under \( \hat{H}^{\prime}(t) \).
    The orange line shows \( \langle \phi(t) | \hat{H}_{\mathrm{P}}^{\mathrm{eff}} | \phi(t) \rangle \), where the initial state \( \ket{\phi(0)} \) is the ground state of \( \hat{H}_{\mathrm{D}}^{\mathrm{eff}} \), and the system evolves under \( \hat{H}^{\mathrm{eff}}(t) \).  
    The green line represents the expectation value of \( \hat{H}_{\mathrm{P}}^{\mathrm{eff}} \) in the ground state of \( \hat{H}_{\mathrm{P}}^{\mathrm{eff}} \), obtained via numerical diagonalization.  
    It can be seen that, when the total annealing time \( T \) is sufficiently long, the expectation value of \( \hat{H}_{\mathrm{P}}^{\mathrm{eff}} \) obtained by solving the Schrödinger equation converges to the ground-state value obtained by numerical diagonalization.  
    In addition, the blue curve exhibits oscillations, and its minimum values are nearly consistent with the orange curve.}
    \label{fig:0323_QIT_T_Hp.pdf}
\end{figure}

\begin{figure}[h!t]
    \centering
    \includegraphics[width=\linewidth]{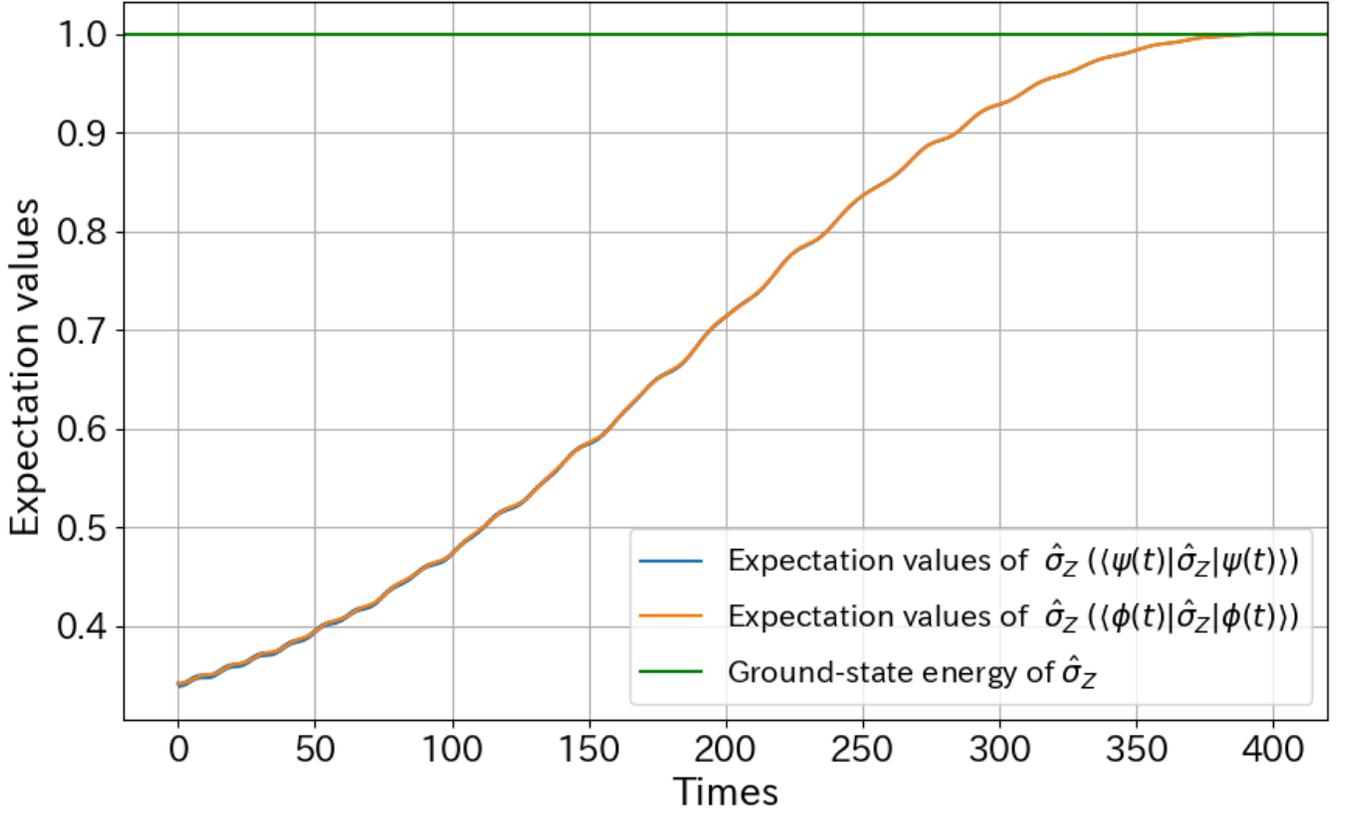}
    \caption{Result of solving the Schrödinger equation for a system in which a single resonator is coupled to a single qubit.  
    The expectation value of \( \hat{\sigma}_z \) at each time \( t \) is plotted.  
    The parameters used are the same as those in Fig.~\ref{fig:0323_QIT_T_Hp.pdf}.  
    The vertical axis represents the expectation value of \( \hat{\sigma}_z \), and the horizontal axis represents time \( t \).  
    The blue line shows \( \langle \psi(t) | \hat{\sigma}_z | \psi(t) \rangle \), where the initial state \( \ket{\psi(0)} \) is the ground state of \( \hat{H}_{\mathrm{D}}^{\mathrm{eff}} \), and time evolution is governed by \( \hat{H}^{\prime}(t) \).  
    The orange line shows \( \langle \phi(t) | \hat{\sigma}_z | \phi(t) \rangle \), where the initial state \( \ket{\phi(0)} \) is also the ground state of \( \hat{H}_{\mathrm{D}}^{\mathrm{eff}} \), and time evolution is governed by \( \hat{H}^{\mathrm{eff}}(t) \).  
    The green line represents the expectation value of \( \hat{\sigma}_z \) in the ground state of \( \hat{H}_{\mathrm{P}}^{\mathrm{eff}} \), obtained via numerical diagonalization.  
    It can be observed that when the total annealing time \( T \) is sufficiently long, the expectation value of \( \hat{\sigma}_z \) obtained from solving the Schrödinger equation converges to the value obtained via numerical diagonalization.  
    Additionally, the blue and orange curves are nearly identical.
    }
    \label{fig:0323_QIT_T_sz.pdf}
\end{figure}

\begin{figure}[h!t]
    \centering
    \includegraphics[width=\linewidth]{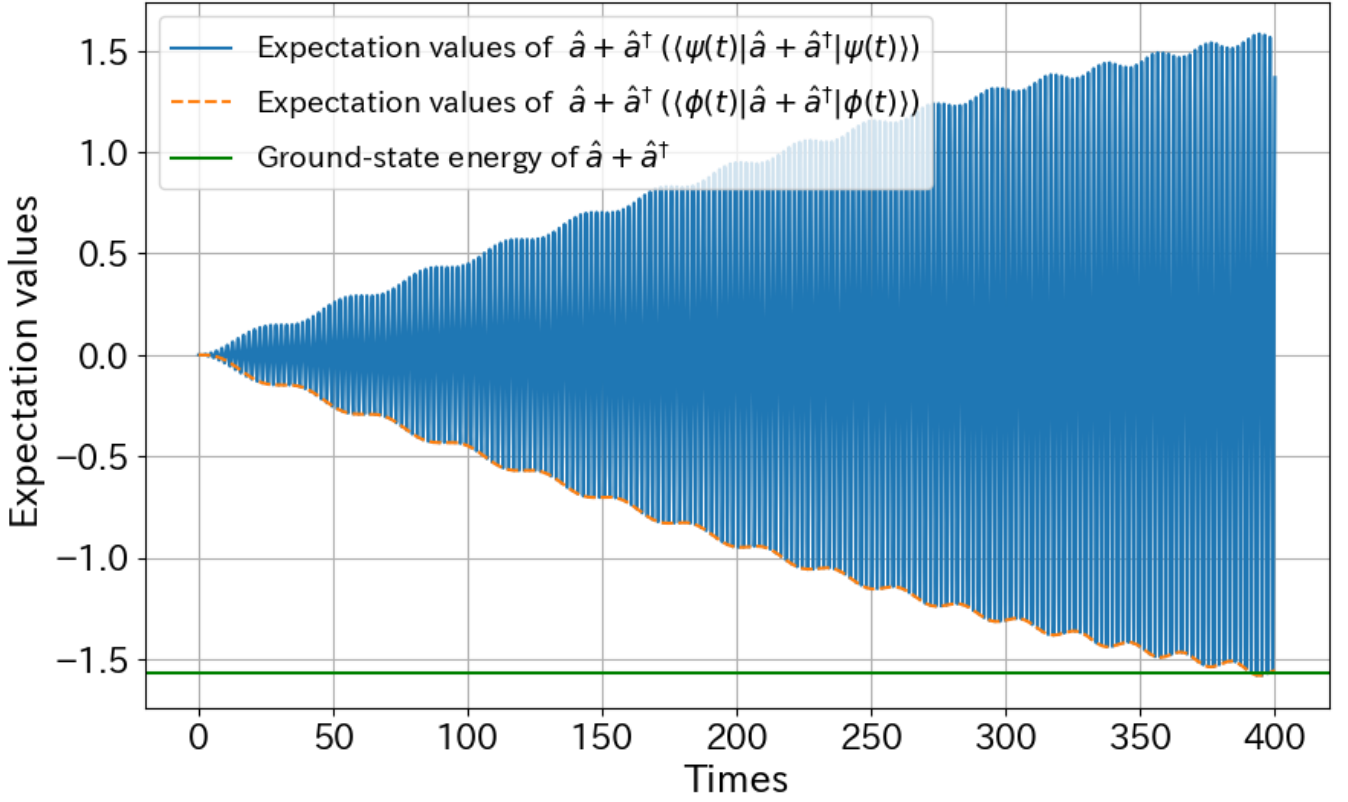}
    \caption{Result of solving the Schrödinger equation for a system in which a single resonator is coupled to a single qubit.  
    The expectation value of \( \hat{a} + \hat{a}^{\dagger} \) at each time \( t \) is plotted.  
    The parameters used are the same as those in Fig.~\ref{fig:0323_QIT_T_Hp.pdf}.  
    The vertical axis represents the expectation value of \( \hat{a} + \hat{a}^{\dagger} \), and the horizontal axis represents time \( t \).  
    The blue line shows \( \langle \psi(t) | (\hat{a} + \hat{a}^{\dagger}) | \psi(t) \rangle \), where the initial state \( \ket{\psi(0)} \) is the ground state of \( \hat{H}_{\mathrm{D}}^{\mathrm{eff}} \), and the system evolves under \( \hat{H}^{\prime}(t) \).  
    The orange line shows \( \langle \phi(t) | (\hat{a} + \hat{a}^{\dagger}) | \phi(t) \rangle \), where the initial state \( \ket{\phi(0)} \) is the ground state of \( \hat{H}_{\mathrm{D}}^{\mathrm{eff}} \), and the system evolves under \( \hat{H}^{\mathrm{eff}}(t) \).  
    The green line represents the expectation value of \( \hat{a} + \hat{a}^{\dagger} \) in the ground state of \( \hat{H}_{\mathrm{P}}^{\mathrm{eff}} \), obtained via numerical diagonalization.  
    It can be seen that, when the total annealing time \( T \) is sufficiently long, the expectation value of \( \hat{a} + \hat{a}^{\dagger} \) obtained by solving the Schrödinger equation converges to the ground-state value obtained by numerical diagonalization.  
    In addition, the blue curve exhibits oscillations, and its minimum values are nearly consistent with the orange curve.
    }
    \label{fig:0323_QIT_T_a+adagg.pdf}
\end{figure}

\begin{figure}[h!t]
    \centering
    \includegraphics[width=\linewidth]{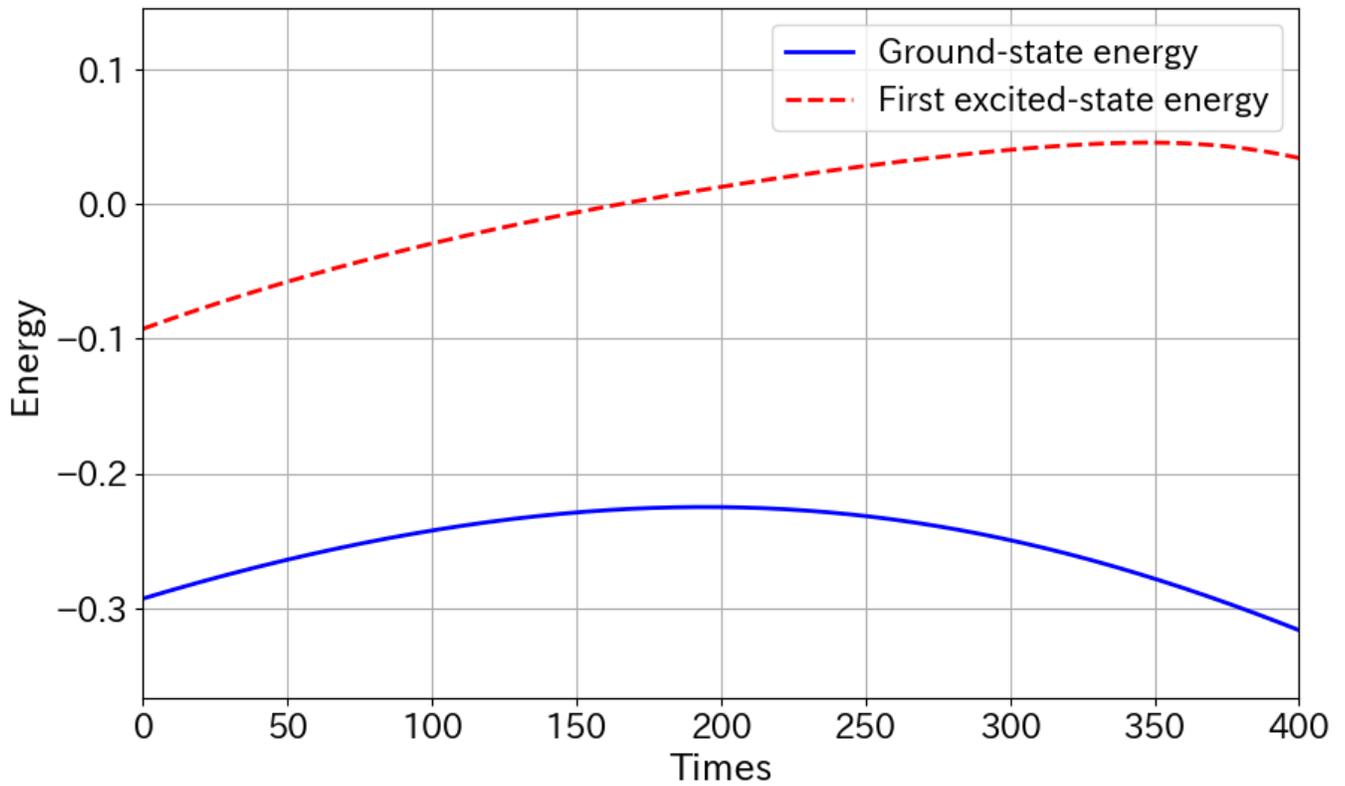}
    \caption{Energy diagram of quantum annealing for a {coupled} system with one resonator and one qubit.  
    The ground-state and first-excited-state energies of the Hamiltonian in Eq.~\eqref{heff(t)} are plotted at each time step.  
    The parameters used are the same as those in Fig.~\ref{fig:0323_QIT_T_Hp.pdf}.  
    The vertical axis represents the {lowest eigenenergy and the second lowest eigenenergy of the effective Hamiltonian.} 
    The horizontal axis represents time \( t \).  
    The blue line shows the ground-state energy, while the red line shows the first-excited-state energy.  
    A finite energy gap between the ground state and the first excited state can be observed.
    }
    \label{fig:0323_QIT_HP_energy_diagram.pdf}
\end{figure}

{Let us investigate the performance of the hybrid quantum annealing with oscillating magnetic fields.}
By numerically solving the Schrödinger equation, we calculated the quantum annealing dynamics of a system in which a single resonator is coupled to a single qubit {under the effect of oscillating magnetic fields}.
The problem Hamiltonian used was given by Eq.\eqref{ippankei.7}, the driver Hamiltonian by Eq.\eqref{ippankei.8}, and the total Hamiltonian by Eq.~\eqref{heff(t)}.
The parameters were set as follows:$L,M=1,\; J,\tilde{J}=0,\;\omega= 153.9,\;h= 153.7,\;\omega_{\mathrm
c}= 154.1,\; B=0.55,\;\lambda=0.30,\; g=0.15,\;\tilde{g}=0.25$.
We plot the expectation value \(\bra{\psi (t)}\hat{H}_{\mathrm{P}}^{\mathrm{eff}}\ket{\psi (t)}\) (blue line), obtained by solving the Schrödinger equation with time evolution under \(\hat{H}(t)\), where the initial state \(\ket{\psi (0)}\) is the ground state of \(\hat{H}_{\mathrm{D}}^{\mathrm{eff}}\).
{Also, we plot} the expectation value \(\bra{\phi (t)}\hat{H}_{\mathrm{P}}^{\mathrm{eff}}\ket{\phi (t)}\) (orange line), obtained by solving the Schrödinger equation under \(\hat{H}^{\mathrm{eff}}(t)\) with initial state \(\ket{\phi (0)}\), also taken as the ground state of \(\hat{H}_{\mathrm{D}}^{\mathrm{eff}}\), at each time $t$ (see Fig.~\ref{fig:0323_QIT_T_Hp.pdf}).
Similarly, Fig.~\ref{fig:0323_QIT_T_sz.pdf} shows the time evolution of the expectation value of $\hat{\sigma}_z$, and Fig.~\ref{fig:0323_QIT_T_a+adagg.pdf} shows the time evolution of the expectation value of $\hat{a}+\hat{a}^{\dagger}$.
{Also, we plot the energy diagram of the effective Hamiltonian in Fig.~\ref{fig:0323_QIT_HP_energy_diagram.pdf}.}

From Fig.~\ref{fig:0323_QIT_T_Hp.pdf}, we observe that the expectation value \(\langle \phi (t)|\hat{H}_{\mathrm{P}}^{\mathrm{eff}}|\phi (t)\rangle \) (orange line), obtained by solving the Schrödinger equation under $\hat{H}^{\mathrm{eff}}$ , converges to the ground-state energy (green line) of \(\hat{H}_{\mathrm{P}}^{\mathrm{eff}}\) obtained via numerical diagonalization.
In contrast, the expectation value \(\langle \psi (t)|\hat{H}_{\mathrm{P}}^{\mathrm{eff}}|\psi (t)\rangle \) (blue line), calculated by solving the Schrödinger equation under $\hat{H}$ , exhibits oscillatory behavior. For sufficiently long times, the {local} minima of this oscillation match the ground-state energy (green line) of \(\hat{H}_{\mathrm{P}}^{\mathrm{eff}}\) obtained via numerical diagonalization.
Furthermore, the  points
{of the local minimum} of the oscillation in the expectation value \(\langle \psi (t)|\hat{H}_\mathrm{P}^{\mathrm{eff}}|\psi (t)\rangle \) (blue line) coincide with the expectation value \(\langle \phi (t)|\hat{H}_\mathrm{P}^{\mathrm{eff}}|\phi (t)\rangle \) (orange line). The following subsection will explain and verify the reason for this behavior.
From
Fig.~\ref{fig:0323_QIT_T_sz.pdf} and 
Fig.~\ref{fig:0323_QIT_T_a+adagg.pdf}, we observe similar behavior for the other observables such as 
$\hat{\sigma}_z$
and $\hat{a} + \hat{a}^{\dagger}$.
Also,
from Fig.~\ref{fig:0323_QIT_HP_energy_diagram.pdf}, we can see that there is a finite energy difference between the ground state energy and the first excited state energy at each time in Eq.~\eqref{heff(t)}.

\clearpage
\subsection{Interpretation of Numerical Results}

In Fig.~\ref{fig:0323_QIT_T_Hp.pdf}, the time evolution of the initial state—prepared as the ground state of \(\hat{H}_{\mathrm{D}}^{\mathrm{eff}}\) —is calculated by solving the Schrödinger equation under both \(\hat{H}^{\prime}(t)\) and \(\hat{H}^{\mathrm{eff}}(t)\), and the expectation value of \(\hat{H}_{\mathrm{P}}^{\mathrm{eff}}\) is plotted at each time $t$.
In the following, we 
{explain the reason why}
these two expectation values coincide at times when \(\omega t=2n\pi\), where $n$ is an integer.

First, we consider the expectation value of \(\hat{H}_{\mathrm{P}}^{\mathrm{eff}}\)at each time $t$, where the time evolution is governed by the Schrödinger equation under \(\hat{H}^{\prime}(t)\) . Specifically, the initial state is prepared as the ground state of \(\hat{H}_{\mathrm{D}}^{\mathrm{eff}}\) , denoted by \( |\psi (t=0)\rangle\).
The time evolution of this state is obtained by solving the Schrödinger equation in the laboratory frame
{such as}
\( i\hbar\frac{\partial |\psi (t)\rangle }{\partial t}=\hat{H}^{\prime}(t)|\psi (t)\rangle\)
Using the evolved state \( |\psi (t)\rangle\), the expectation value of \(\hat{H}_{\mathrm{P}}^{\mathrm{eff}}\) can be computed as\(\langle \psi (t)|\hat{H}_{\mathrm{P}}^{\mathrm{eff}}|\psi (t)\rangle\).
\begin{equation}
    \langle \psi (t)|\hat{H}_{\mathrm{P}}^{\mathrm{eff}}|\psi (t)\rangle
    \label{setumei1.1}
\end{equation}

On the other hand, we consider the expectation value of \(\hat{H}_{\mathrm{P}}^{\mathrm{eff}}\) at each time $t$, where the time evolution is governed by the Schrödinger equation under \(\hat{H}^{\mathrm{eff}}(t)\). Specifically, the initial state is prepared as the ground state of \(\hat{H}_{\mathrm{D}}^{\mathrm{eff}}\) , denoted by \( |\phi (t=0)\rangle\). The time evolution is determined by solving the Schrödinger equation in the rotating frame: \( i\hbar\frac{\partial |\phi (t)\rangle }{\partial t}=\hat{H}^{\mathrm{eff}}|\phi (t)\rangle\) and the corresponding expectation value is calculated as
\begin{equation}
    \langle \phi (t)|\hat{H}_{\mathrm{P}}^{\mathrm{eff}}|\phi (t)\rangle
    \label{setumei1.2}
\end{equation}

Here, we assume the relation \(|\phi (t)\rangle=\hat{U}|\psi (t)\rangle\).Under this assumption, Eq.~\eqref{setumei1.2} can be rewritten as follows:
\begin{equation}
    \langle\phi (t)|\hat{H}_{\mathrm{P}}^{\mathrm{eff}}|\phi (t)\rangle=\langle\psi (t)|\hat{U}^\dagger\hat{H}_{\mathrm{P}}^{\mathrm{eff}}\hat{U}|\psi (t)\rangle
    \label{setumei1.3}
\end{equation}

Next, we consider the transformation \(\hat{U}^\dagger\hat{H}_{\mathrm{P}}^{\mathrm{eff}}\hat{U}\) , which yields
\begin{equation}
    \begin{aligned}
    \hat{U}^\dagger\hat{H}_{\mathrm{P}}^{\mathrm{eff}}\hat{U}&=\hat{U}^\dagger\Big(\dfrac{h-\omega}{2}\hat{\sigma_{z}}
    +
    g\hat{\sigma_z}(\hat{a} + \hat{a}^\dagger)
    +
    \frac{\tilde{g}}{2}\hat{\sigma_z}\hat{a}^\dagger\hat{a}
    +
    (\omega_{\mathrm{c}}-\omega)\hat{a}^\dagger\hat{a}
    +
    \frac{\lambda}{2}(\hat{a} + \hat{a}^\dagger)
    \Big)\hat{U}
    \\
    &=\dfrac{ h-\omega}{2}\hat{\sigma_{z}}
    +
    \frac{g}{2}\hat{\sigma_z}
    \left(
    e^{i\omega t}\hat{a} + e^{-i\omega t}\hat{a}^\dagger
    \right)
    +
    \tilde{g}\hat{\sigma_z}\hat{a}^\dagger\hat{a}
    +
    (\omega_{c}-\omega)(\hat{a}^\dagger\hat{a})
    +
    \frac{\lambda}{2}\left(
    e^{i\omega t}\hat{a} + e^{-i\omega t}\hat{a}^\dagger
    \right)
    \label{setumei1.4}   
    \end{aligned}
\end{equation}
and we see that Eq.~\eqref{setumei1.4} coincides with \(\hat{H}_{\mathrm{P}}^{\mathrm{eff}}\) when \(\omega t=2n\pi\), where $n$ is an integer.
\begin{equation}
    \begin{aligned}
    \hat{U}^\dagger\hat{H}_{\mathrm{P}}^{\mathrm{eff}}\hat{U}&=\dfrac{h-\omega}{2}\hat{\sigma_{z}}+
    (\omega_{\mathrm{c}}-\omega)\hat{a}^\dagger\hat{a}
    +
    \frac{\lambda}{2}(\hat{a} + \hat{a}^\dagger)
    +
    g\hat{\sigma_z}(\hat{a} + \hat{a}^\dagger)
    +
    \frac{\tilde{g}}{2}\hat{\sigma_z}\hat{a}^\dagger\hat{a}
    \\
    &=\hat{H}_{\mathrm{P}}^{\mathrm{eff}}
    \label{setumei1.5}   
    \end{aligned}
\end{equation}
Therefore, Eq.\eqref{setumei1.3} becomes
\begin{equation}
    \begin{aligned}
     \langle\phi (t)|\hat{H}_{\mathrm{P}}^{\mathrm{eff}}|\phi (t)\rangle&=\langle\psi (t)|\hat{U}^\dagger\hat{H}_{\mathrm{P}}^{\mathrm{eff}}\hat{U}|\psi (t)\rangle
     \\
     &=\langle\psi (t)|\hat{H}_{\mathrm{P}}^{\mathrm{eff}}|\psi (t)\rangle
    \label{setumei1.6}   
    \end{aligned}
\end{equation}
which matches Eq.\eqref{setumei1.1}. In the following section, we numerically verify whether this equality also holds in practice.

\begin{figure}[H]
    \centering
    \includegraphics[width=\linewidth]{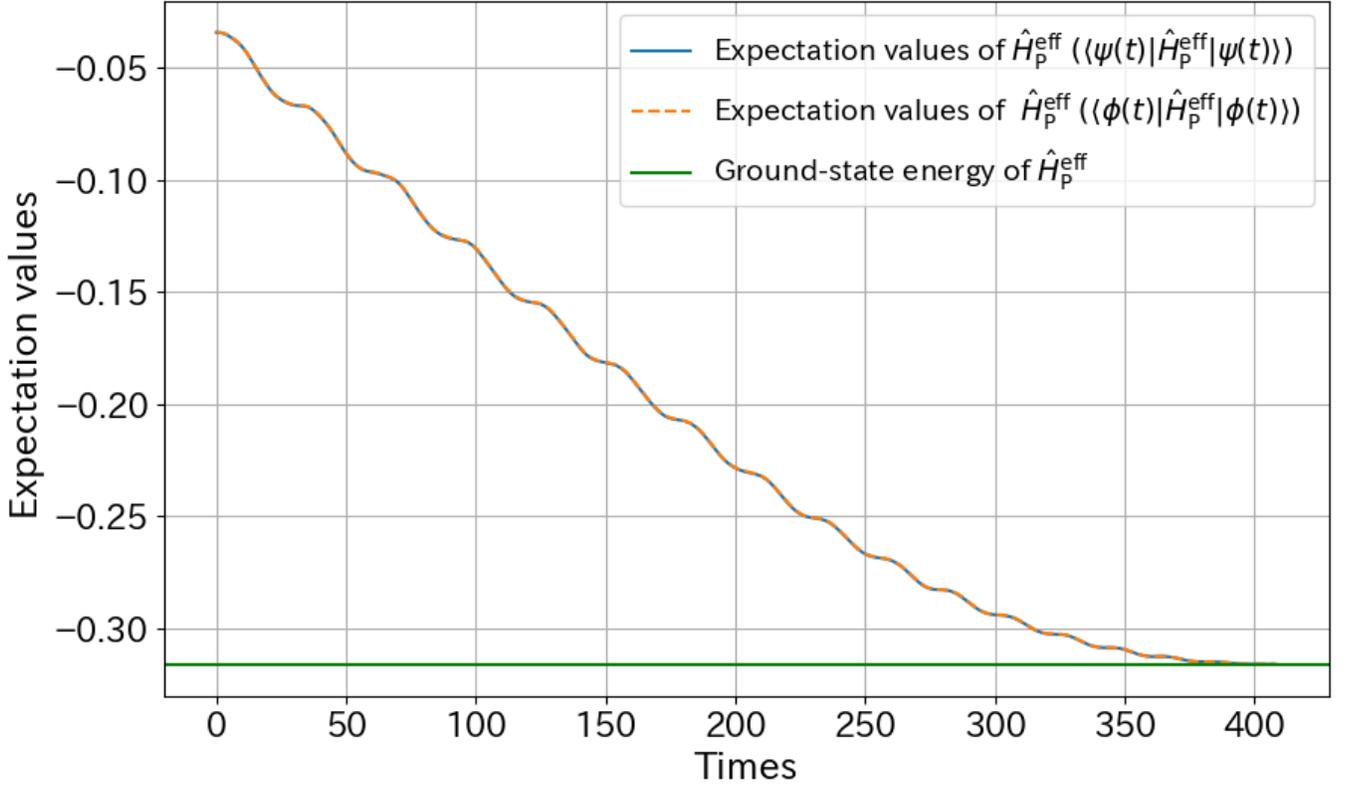}
    \caption{Result of solving the Schrödinger equation for a system in which a single resonator is coupled to a single qubit.  
    The expectation value of \( \hat{H}_{\mathrm{P}}^{\mathrm{eff}} \) is plotted at each time step.  
    The parameters used are the same as in Fig.~\ref{fig:0323_QIT_T_Hp.pdf}, except that the total annealing time was set to \( T \simeq 408.2 \) (corresponding to \( \omega t = 2n\pi \)).  
    The vertical axis represents the expectation value of \( \hat{H}_{\mathrm{P}}^{\mathrm{eff}} \), and the horizontal axis represents time \( t \).  
    The blue line shows \( \langle \psi(t) | \hat{H}_{\mathrm{P}}^{\mathrm{eff}} | \psi(t) \rangle \), where \( \ket{\psi(0)} \) is the ground state of \( \hat{H}_{\mathrm{D}}^{\mathrm{eff}} \) and the system evolves under \( \hat{H}^{\prime}(t) \).  
    The orange line shows \( \langle \phi(t) | \hat{H}_{\mathrm{P}}^{\mathrm{eff}} | \phi(t) \rangle \), where \( \ket{\phi(0)} \) is also the ground state of \( \hat{H}_{\mathrm{D}}^{\mathrm{eff}} \), and the system evolves under \( \hat{H}^{\mathrm{eff}}(t) \).  
    The green line represents the ground-state energy of \( \hat{H}_{\mathrm{P}}^{\mathrm{eff}} \), obtained via numerical diagonalization.  
    It can be observed that, with sufficiently long total annealing time \( T \), the expectation value of \( \hat{H}_{\mathrm{P}}^{\mathrm{eff}} \) obtained by solving the Schrödinger equation converges to the ground-state energy from numerical diagonalization.  
    Furthermore, the blue and orange lines are in excellent agreement.
    }
    \label{fig:0323_QIT_T_Hp_pi}
\end{figure}

Using the same parameters as in Section~\ref{sec:a+adag3}, we performed quantum annealing and numerically solved the Schrödinger equation only at times \(t=2n\pi/{\omega}\). The results are plotted in Fig.~\ref{fig:0323_QIT_T_Hp_pi}. 
{Here,}
it is observed that the expectation value \(\bra{\psi (t)}\hat{H}_{\mathrm{P}}^{\mathrm{eff}}\ket{\psi (t)}\), obtained by solving the Schrödinger equation under \(\hat{H}^{\prime}(t)\) with the initial state \(\ket{\psi (0)}\) being the ground state of \(\hat{H}_{\mathrm{D}}^{\mathrm{eff}}\) , coincides with the expectation value \(\bra{\phi (t)}\hat{H}_{\mathrm{P}}^{\mathrm{eff}}\ket{\phi (t)}\), obtained by solving the Schrödinger equation under \(\hat{H}^{\mathrm{eff}}(t)\) with the initial state \(\ket{\psi (0)}\) also being the ground state of \(\hat{H}_{\mathrm{D}}^{\mathrm{eff}}\).
Based on the above, we confirm that the result obtained by solving the Schrödinger equation with the Hamiltonian in the rotating frame using the rotating wave approximation
coincides with that obtained using the Hamiltonian in the laboratory frame {if we consider the local minimum of the oscillations}. 

\bibliography{AQMbib}

\end{document}